# ARE THE LYMAN ALPHA FOREST "CLOUDS" EXPANDING PANCAKES?

*Some theoretical implications of the recent size determinations of Ly$\alpha$ absorbers*

M.G. HAEHNELT

*Max-Planck-Institut für Astrophysik*
*Karl-Schwarzschild-Straße 1*
*85740 Garching, Germany*

**Abstract.** The large sizes of Ly$\alpha$ "clouds" inferred from coincident absorption in the spectrum of close quasar pairs suggests that these are transient flattened structures of small overdensity. It is argued that the observed absorbers should be preferentially located in underdense regions of the universe and should typically expand faster than the Hubble flow.

## 1. Introduction

The use of quasar absorption spectra to study the distribution of neutral hydrogen at high redshift is a well established cosmological tool. However, the physical nature of Ly$\alpha$ forest absorbers itself had long remained unclear (see also the contribution by Michael Rauch, these proceedings). This is mainly due to the fact that basic properties like typical size, density and mass of the absorbers sucessfully eluded meaningful observational constraints. This has changed with the recent detection of coincident absorption lines in the two close quasar pairs Q0107-025AB and Q1343+266AB with proper separations of $360h^{-1}$ kpc and $40h^{-1}$ kpc at redshifts $z \sim 1$ and $z \sim 2$ (Dinshaw et al. 1994, Dinshaw et al. 1995, Bechthold et al. 1994). The observed fraction of coincident lines of about 50-80% implies that the absorbing structures coherently cover (with covering factor close to unity) an area which is up to a Mpc across. The inferred size depends somewhat on the shape and internal column density distribution of the absorbing structure and also on the cosmological parameter, but is in any case considerably larger than predicted by most of the models discussed earlier.

## 2. Fiducial parameter of Ly$\alpha$ forest absorbers

Knowing the transverse size of the absorbers renders it possible to infer a variety of otherwise poorly constrained physical parameters. The typical neutral hydrogen density is $n_{\rm HI} \sim 1.6 \times 10^{-10}\, h\, f_1^{-1}\, N_{14}\, r_{100}^{-1}\,{\rm cm}^{-3}$, where $N_{14}$ and $r_{100}$ are the column density of neutral hydrogen and the transverse "radius" of the absorber scaled to $10^{14}\,{\rm cm}^{-2}$ and $100 h^{-1}$ kpc. There is still some uncertainty left parameterized as $f_1$ which is due to the a priori unknown ratio between the measured transverse size of the absorber and its extent along the line of sight. It has been known for a long time that the absorbers must be highly ionized by the UV background. Taking the fiducial value for its intensity $I_\nu = I_{21} \times 10^{-21}\,{\rm erg\,s^{-1}\,cm^{-2}\,Hz^{-1}\,sr^{-1}}$ (Bechtold 1994) gives a typical neutral hydrogen fraction of $x \sim 4 \times 10^{-6} h^{1/2} f_1^{-0.5} I_{21}^{-0.5} N_{14}^{0.5} r_{100}^{-0.5}$. This corresponds to a typical total hydrogen density and baryonic mass of $n_{\rm H} \sim 4 \times 10^{-5}\, h^{1/2}\, f_1^{-0.5}\, I_{21}^{0.5}\, N_{14}^{0.5}\, r_{100}^{-0.5}\,{\rm cm}^{-3}$ and $M_{\rm bar} \sim 8 \times 10^9\, h^{-5/2}\, f_1^{0.5}\, I_{21}^{0.5}\, N_{14}^{0.5}\, r_{100}^{2.5}\, M_\odot$. We can further use the observed column density distribution $f(N)$ to obtain an estimate of the overall fraction of the critical density contained in the Ly$\alpha$ forest

$$\Omega_{\rm Ly\alpha} = \frac{\mu\, m_H\, H_0}{c\, \rho_{0crit}} \int x^{-1}(N)\, N\, f(N)\, {\rm d}N \sim 0.08\, h^{-3/2}\, f_1^{0.5}\, I_{21}^{0.5}\, r_{100}^{0.5}, \qquad (1)$$

where $\mu m_H$ is the mean mass per hydrogen atom and the other symbols have their usual meaning. Equation (1) shows that the recent size estimates make the inferred $\Omega_{\rm Ly\alpha}$ uncomfortably large in comparison to the nucleosynthesis constraint on the baryonic matter content of the universe. As argued by Rauch & Haehnelt (1995) this suggests a flattened geometry of the absorbers with $f_1 \lesssim 0.1$.

## 3. What is the nature of the Ly$\alpha$ forest absorbers?

The main properties of the absorbers can be summarized as follows:

- A significant fraction of all baryons (of order unity) is contained in them.
- The baryonic mass of the individual absorber is similar to that of a $L_*$ galaxy, but the inferred number density exceeds this of $L_*$ galaxies by a factor of about $10 - 30$.
- They are overdense compared to the mean baryonic density by a factor of about a few to ten.
- They are likely to have a flattened geometry.

All of these points strongly argue against them being virialized objects. As realized by Cen et al. (1994) and others the Ly$\alpha$ forest absorbers are most likely some modest transient density fluctuations of the intergalactic medium caused by the large scale flows and density fluctuations of the dynamically dominant dark matter component of the universe.

However, the question remains what sort of underlying structures are causing these density fluctuations and in which dynamical state they are. The standard paradigm for the origin of large-scale structure is the growth of small primordial density fluctuations which can be described by a Gaussian random field. In the following I will imagine the universe to be divided into regions of equal mass and use the Zeldovich approximation (Zeldovich 1970) to describe the dynamical evolution of such regions. The Zeldovich approximation is an astonishingly good description even in the mildly nonlinear regime and should give a good qualitative understanding. The trajectory of a particle in an Einstein-de-Sitter universe is then given by

$$r_i(\mathbf{q}, t) = a(t) \left[ q_i + (a(t) - 1) \right] \Phi_{,i}(\mathbf{q}), \tag{2}$$

where $a(t)$ is the global growth factor, $\mathbf{q}$ is the initial Lagrangian coordinate and $\Phi$ is the gravitational potential. The distribution of the eigenvalues of the Zeldovich tensor $\Phi_{,ij}$ was first derived by Doroskhevich (1974) and can be written as

$$p(\lambda_1, \lambda_2, \lambda_3) = \frac{15^3}{8\pi\sqrt{5}} (\lambda_3 - \lambda_2)(\lambda_3 - \lambda_1)(\lambda_2 - \lambda_1) \tag{3}$$

$$\times \exp\left\{ -\frac{3}{2} [2(\lambda_1^2 + \lambda_2^2 + \lambda_3^2) - (\lambda_1\lambda_2 + \lambda_1\lambda_3 + \lambda_2\lambda_3)] \right\} d^3\lambda, \tag{4}$$

where $\lambda_1 \leq \lambda_2 \leq \lambda_3$ (Steinmetz & Bartelmann 1995). Negative eigenvalues correspond to a contraction along the corresponding axis with respect to the Hubble flow, while positve eigenvalues describe expansion. The density contrast in the linear regime is determined by the sum of all eigenvalues $\delta = (\rho - \langle \rho \rangle)/\langle \rho \rangle = -\sigma(\lambda_1 + \lambda_2 + \lambda_3)$ and the rms amplitude of the density fluctuations $\sigma$. Completely collapsed and virialized objects generally originate from regions, where $\delta$ is initially positive. However, collapse along one axis should be sufficient to produce low-column density absorption lines even if the other two axis will expand for ever. This is considerably more likely to occur than a full collapse along all three axis. Furthermore an object collapsed along one axis only (a pancake) has a much larger cross section than one which is collapsed along more than one axis (a filament or fully virialized object). The solid curve in Fig.1a shows the fraction of the mass in regions where collapse along one axis has occured according to the Zeldovich approximation ($\lambda_1 \sigma < -1$) while the other two eigenvalues are positive. For the dashed curve $(\lambda_1 + \lambda_2 + \lambda_3) > 0$ was imposed as second constraint. Both fractions are about 50 % and depend only weakly on $\sigma$.

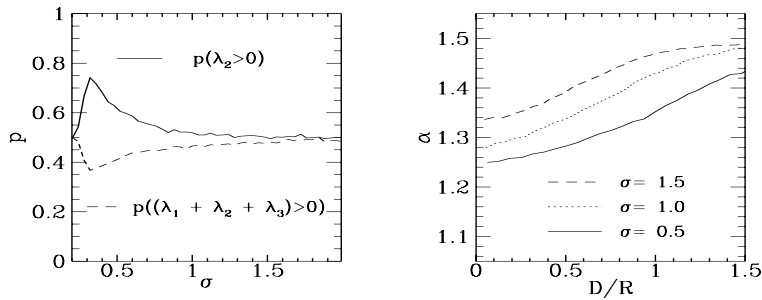

Figure 1. The solid curve of the left diagram shows the fraction of the mass in regions where collapse along one axis has occured according to the Zeldovich approximation ($\lambda_1 \sigma < -1$) while the other two eigenvalues are positive. For the dashed curve ($\lambda_1 + \lambda_2 + \lambda_3) > 0$ was imposed as second constraint. The right diagram shows the ratio $\alpha$ between the expansion velocity in the plane of a randomly orientated Zeldovich pancake giving rise to coincident absorption as a function of the ratio between line-of-sight separation and radius of the pancake. The three curves are for different values of the rms amplitude of the density fluctuations $\sigma$ as indicated in the plot.

To estimate the dynamical state of a typical absorber we can use again the Zeldovich approximation. The velocity between two points in the plane of the pancake will then be linearly proportional to their distance (just as in Hubble's law). Figure 1b shows the ratio $\alpha$ between the expansion velocity in the plane of the pancake and the Hubble velocity for randomly orientated pancakes. The typical pancake giving rise to coincident absorption is expanding about 30% faster than the Hubble flow. This is due to the fact that positive and negative eigenvalues of the Zeldovich tensor are equally likely and that the pancakes expanding fastest have the largest cross section for absorption. For the same reason low-column-density absorption lines should be preferentially embedded in underdense region of the universe.

## 4. Observational tests

The model described above predicts rather small velocity differences between coincident absorption lines in adjacent lines of sight

$$\Delta v = \alpha \cot\theta \, H(z) \, D \sim 50 \, \alpha \, \cot\theta \left(\frac{1+z}{3}\right)^{3/2} \left(\frac{D}{100 h^{-1} \, \text{kpc}}\right) \text{ km s}^{-1} \quad (5)$$

where $\theta$ is the angle between the plane of the pancake and the line of sight and D is the proper distance between the two lines of sight. The probability distribution of the orientation angle of Zeldovich-pancakes giving rise to coincident absorption is plotted in Fig.2. Face-on pancakes are more likely to be responsible for coincident absorption. This is mainly due to their larger cross section. The observational results are so far inconclusive. The velocity differences observed in the spectra of Q0107-025AB and Q1343+266AB are smaller than would be expected for virialized objects of this size, but the quoted values are hardly larger than the errors. Furthermore, the line-of sight differences are uncomfortably small and the column densities of the coincident lines rather large for this kind of test. However, the model should become testable in the near future with quasar pairs of somewhat wider separation and improved signal-to-noise.

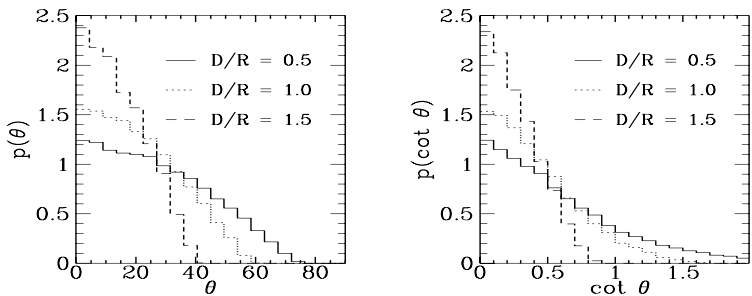

*Figure 2.* The left diagram shows the probability distribution of the orientation angle $\theta$ between the plane of a Zeldovich-pancake giving rise to coincident absorption and the line of sight. The three curves are for different values of the ratio between line-of-sight separation and radius of the pancake as indicated in the plot. The right diagram shows the probability distribution of $\cot\theta$.

## 5. Conclusions

The recent measurement of the sizes of Ly$\alpha$ forest absorbers has changed our understanding of their nature. They are now believed to be transient density fluctuations of the intergalactic medium reflecting the evolution of pancake-like structures in the dark-matter component of the universe. These pancakes are typically expanding about 30% faster than the Hubble flow and in most cases they will never collapse, but rather be incorporated into larger structures.